\documentclass[preprint,aps,nofootinbib,superscriptaddress]{revtex4}
\usepackage{amssymb}
\usepackage{amsmath}
\usepackage{graphicx}

\newcommand\beq{ \begin{eqnarray} }
\newcommand\eeq{ \end{eqnarray} }

\begin{document}

\title{On the Quark Mass Dependence of Two Nucleon Observables}
\author{Jiunn-Wei Chen}
\email{jwc@phys.ntu.edu.tw}
\affiliation{Department of Physics and Center for Theoretical Sciences, National Taiwan
University, Taipei 10617, Taiwan}
\author{Tze-Kei Lee }
\email{chikei@gmail.com}
\affiliation{Department of Physics and Center for Theoretical Sciences, National Taiwan
University, Taipei 10617, Taiwan}
\author{C.-P. Liu}
\email{cpliu@mail.ndhu.edu.tw}
\affiliation{Department of Physics, National Dong Hwa University, Hua-Lien 974, Taiwan}
\author{Yu-Sheng Liu}
\email{mestelqure@gmail.com}
\affiliation{Department of Physics and Center for Theoretical Sciences, National Taiwan
University, Taipei 10617, Taiwan}

\begin{abstract}
We study the implications of lattice QCD determinations of the S-wave
nucleon-nucleon scattering lengths at unphysical light quark masses. It is
found that with the help of nuclear effective field theory (NEFT), not only
the quark mass dependence of the effective range parameters, but also the
leading quark mass dependence of all the low energy deuteron matrix elements
can be obtained. The quark mass dependence of deuteron charge radius,
magnetic moment, polarizability and the deuteron photodisintegration cross
section are shown based on the NPLQCD lattice calculation of the scattering
lengths at 354 MeV pion mass and the NEFT power counting scheme of Beane,
Kaplan and Vuorinen. Further improvement can be obtained by performing the
lattice calculation at smaller quark masses. Our result can be used to
constrain the time variation of isoscalar combination of $u$ and $d$ quark
mass $m_{q}$, to help the anthropic principle study to find the $m_{q}$
range which allows the existence of life, and to provide a weak test of the
multiverse conjecture.
\end{abstract}

\maketitle

\section{Introduction}

A very interesting aspect of lattice QCD (LQCD) calculations is that one can
study the quark mass dependence of physical observables which are otherwise
hard to measure with experiments. This information could be used to
constrain the time variation \cite{Dirac,Webb:1998cq,Webb:2000mn} of quark
masses in the evolution of the universe \cite%
{Beane:2002vs,Beane:2002xf,Epelbaum:2002gb,Flambaum:2007mj}. It could also
shed light on how finely tuned the quark masses should be \cite%
{Flambaum:2002de,Kneller:2003xf,Kneller:2003ka,Landau:2004rj,Dent:2007zu}\
such that light nuclei can be synthesized through the usual pathway of Big
Bang Nucleosynthesis (BBN) \cite{Alpher:1948ve,Burles:2000zk} and make the
familiar carbon based life forms possible.

Much has been learned from the $u$ and $d$ quark mass (we will work in the
isosymmetric limit $m_{u}=m_{d}=m_{q}$) dependence of the meson and single
baryon observables \cite{Scholz:2009yz,Alexandrou:2010cm}\ through lattice
QCD (LQCD) \cite{Wilson:1974sk}, chiral perturbation theory (ChPT) \cite%
{Gasser:1983yg,Gasser:1984gg,Jenkins:1990jv,Bernard:1995dp} and experimental
data. In principle, lattice QCD can map out all the $m_{q}$ dependence for
these observables. However, most of the calculations are done with $m_{q}$'s
larger than their physical values, because it requires more computing
resources to work with smaller $m_{q}$. Fortunately, ChPT, which is an
effective field theory (EFT) of QCD, can be used to described the $m_{q}$
dependence once the unknown parameters in the theory are fixed by either
experiments or lattice data.

In the multi-baryon sector, much progress has been made in LQCD in two
nucleon \cite%
{Fukugita:1994ve,Beane:2006mx,Ishii:2006ec,Aoki:2008hh,Beane:2009py},
nucleon-hyperon \cite{Beane:2006gf}, triton \cite%
{Beane:2009gs,Yamazaki:2009ua} and $\alpha $-particle \cite{Yamazaki:2009ua}
systems (see \cite{Savage:2010cx} for a brief review). However, for two
nucleon systems, so far only the S-wave scattering lengths have been
computed with 354 MeV or heavier pion mass $m_{\pi }$. (Note that the
physical pion mass $m_{\pi }^{phys}\simeq 138$ MeV, and there is a
one-to-one correspondence between $m_{\pi }$ and $m_{q}$, e.g. $m_{\pi
}\propto m_{q}^{1/2}$ as $m_{q}\rightarrow 0$. So the $m_{\pi }$ and $m_{q}$
dependence can be converted to each other.) Even so, as will be demonstrated
in this work, this information is enough to determine the leading $m_{\pi }$
dependence\ of all the low energy matrix elements involving deuterons.

We will focus on processes with the typical momentum $p\ll m_{\pi }$, such
that the pions can be taken as heavy particles and integrated out of the
theory. This theory is known as pionless theory \cite%
{Kaplan:1996nv,Bedaque:1997qi,Chen:1999tn,Beane:2000fi}. The information of
the pion dynamics in the pionful theory is now encoded in the $m_{\pi }$
dependent couplings of the pionless theory. It is found that, all the
leading $m_{\pi }$ dependence in deuteron matrix elements in the pionless
theory can be computed using the pionful theory together with the $m_{\pi }$
dependence\ of the S-wave scattering lengths obtained from LQCD. Thus, once
they are fixed at $m_{\pi }^{phys}$, their values at other pion masses are
also known.

Of course, one can still work with the pionful theory. The matching is a
convenient but not necessary step to take. One advantage of working with the
theory without pions is that once the $m_{\pi }$ dependence of the couplings
are worked out, one can just perform the calculation in the pionless theory
instead of the more complicated pionful theory. As an explicit example, we
match the pionful theory based on Beane, Kaplan and Vuorinen's (BKV) \cite%
{Beane:2008bt} power counting scheme to a pionless theory. This allows the
matching been done analytically. However, the method can be applied to other
power schemes as well.

\section{Power Counting Schemes in Nuclear Effective Field Theory}

Currently, there are several power counting schemes for the nuclear
effective theory used for multinucleon systems. Power counting means
counting the power of the small expansion parameter of a Feynman diagram,
such that one can organize the computation in a series expansion of this
parameter. In nuclear EFT, the small expansion parameter $Q$ is $m_{\pi
}/\Lambda $ and $p/\Lambda $, where $\Lambda $ is the cut-off scale. Here we
briefly review some popular power counting schemes.

In Weinberg's scheme \cite{Weinberg:1990rz,Weinberg:1991um,Weinberg:1992yk},
power counting is done to the potential of the Lippmann-Schwinger equation,
not the diagram. The leading-order (LO) potential involves the one pion
exchange (OPE) potential and the delta function potential from contact
interactions. Subtracting the infinities in the LO\ diagram requires higher
order operators with high power of quark mass insertions. Thus, the result
has cut-off dependence that cannot be removed \cite%
{Kaplan:1996xu,Beane:2001bc,Nogga:2005hy}. A similar situation happens to
higher partial waves as well \cite{Nogga:2005hy}. However, within a
reasonable range of cut-off, the scheme works well numerically with
impressive fits to nucleon-nucleon (NN) scattering phase shift data at the
fourth order \cite%
{Ordonez:1992xp,Ordonez:1993tn,vanKolck:1994yi,Ordonez:1995rz,Friar:1998zt,Rentmeester:1999vw, Bernard:1996gq,Epelbaum:1999dj,Epelbaum:2000mx,Epelbaum:2002ji,Epelbaum:2003xx,Epelbaum:2004fk, Entem:2001cg,Entem:2002sf,Entem:2003ft, PavonValderrama:2005wv,PavonValderrama:2005uj}%
.

The alternative KSW scheme \cite{Kaplan:1998we,Kaplan:1998tg} counts the
diagrams near the non-trivial UV fixed point of the four-nucleon operators
such that the cut-off dependence is removed and diagrams of the same order
are of equal size. The LO S-wave diagrams only contain non-derivative
four-nucleon contact interactions, while the next-to-leading-order (NLO)
contains OPE diagrams and diagrams with higher order four-nucleon operators.
However, numerically, the convergence is not good in the $^{3}S_{1}$ channel
due to the singular\ nature of the tensor pion exchange potential at short
distance $\widehat{\mathbf{r}}_{i}\widehat{\mathbf{r}}_{j}/r^{3}$, where $%
\mathbf{r}$ is the distance between two nucleons \cite%
{Fleming:1999bs,Fleming:1999ee}. The suggests that the tensor pion exchange
might not be perturbative.

In view of this problem, the tensor pion exchange is resummed at the LO in
the BBSvK scheme \cite{Beane:2001bc} where the $^{1}S_{0}$ channel follows
the KSW power counting while the $^{3}S_{1}$ channel follows the Weinberg's
power counting. It was shown that the cut-off can be removed in this scheme.

The BKV scheme \cite{Beane:2008bt} seeks to fix the same problem by
introducing a Pauli-Villars (PV) field in the $^{3}S_{1}$ channel to remove
the short distance part of the singular tensor potential. The resulting $%
^{3}S_{1}$ phase shift is convergent. The price to pay is that the PV mass $%
\lambda $ is counted as the same order as $m_{\pi }$, but numerically it is
close to the cut-off scale. However, its analytic result is very convenient
to perform the matching to a pionless theory. Thus, we will adopt the BKV
scheme in this work.

\section{The Quark Mass Dependence of Effective Range Parameters}

The S-wave nucleon-nucleon (NN) scattering amplitude is

\begin{equation}
\mathcal{A}=\frac{4\pi }{M}\frac{1}{p\cot \delta -ip},  \label{A}
\end{equation}%
where $M=938.92$ MeV is the nucleon mass, $p$ is the magnitude of the
nucleon three-momentum in the center-of-mass (CM) frame and $\delta $ is the
S-wave phase shift. If the interaction (potential) is localized, then $%
\delta $ has the expansion \cite{Bethe:1949yr,Bethe:1950jm}

\begin{equation}
p\cot \delta =-\frac{1}{a}+\frac{1}{2}r_{0}p^{2}+\ldots ,  \label{delta}
\end{equation}%
where the effective range parameters (ERP's) $a$ and $r_{0}$ are the
scattering length and the effective range, respectively. The shape parameter
and higher order terms are not shown.

In the BKV scheme, the amplitude can be expanded in powers of the small
expansion parameter $Q$%
\begin{equation}
\mathcal{A}=\mathcal{A}_{-1}+\mathcal{A}_{0}+\mathcal{A}_{1}+\ldots ,
\end{equation}%
where $\mathcal{A}_{n}$ is of order $Q^{n}$ in the expansion. Hence

\begin{equation}
p\cot \delta =ip+\frac{4\pi }{M\mathcal{A}_{-1}}-\frac{4\pi \mathcal{A}_{0}}{%
M\mathcal{A}_{-1}^{2}}+\ldots .  \label{eq1}
\end{equation}

\subsection{Working in the BKV scheme as an explicit example}

The BKV scheme is the same as the KSW scheme in the $^{1}S_{0}$ channel but
different in the $^{3}S_{1}$ channel. The leading order (LO) amplitude of
channel $i$ arises from the diagrams in Fig.~5 of Ref. \cite{Kaplan:1998we}

\begin{equation}
\mathcal{A}_{-1}^{(i)}={\frac{-C_{0}^{(i)}}{\left[ 1+{\frac{C_{0}^{(i)}M}{%
4\pi }}(\mu +ip)\right] }}\ \ \ ,  \label{eq2}
\end{equation}%
where $C_{0}$ is the LO four-nucleon non-derivative coupling which is
independent of $m_{q}$. The next-to-leading-order (NLO) amplitude arises
from the diagrams in Fig.~6 of Ref. \cite{Kaplan:1998we} plus the associated
diagrams with the Pauli-Villars fields 
\begin{eqnarray}
\mathcal{A}_{0}^{(^{1}S_{0})} &=&\mathcal{A}_{0,a}^{(^{1}S_{0})}+\mathcal{A}%
_{0,b}^{(^{1}S_{0})}\left( m_{\pi }\right)  \notag \\
\mathcal{A}_{0}^{(^{3}S_{1})} &=&\mathcal{A}_{0,a}^{(^{3}S_{1})}+\mathcal{A}%
_{0,b}^{(^{3}S_{1})}\left( m_{\pi }\right) -\epsilon \mathcal{A}%
_{0,b}^{(^{3}S_{1})}\left( \lambda \right) ,
\end{eqnarray}%
where $\epsilon $ is introduced to keep track of the difference between the
BKV and the KSW power counting. $\epsilon =1$ gives the BKV result while $%
\epsilon =0$ gives the KSW result. 
\begin{align}
\mathcal{A}_{0,a}^{(i)}& ={\frac{-\left(
C_{2}^{(i)}p^{2}+C_{0,0}^{(i)}\right) }{\left[ 1+{\frac{C_{0}^{(i)}M}{4\pi }}%
(\mu +ip)\right] ^{2}}}  \notag \\
\mathcal{A}_{0,b}^{(i)}\left( m_{\pi }\right) & =\frac{-D_{2}^{(i)}m_{\pi
}^{2}}{\left[ 1+{\frac{C_{0}^{(i)}M}{4\pi }}(\mu +ip)\right] ^{2}}
\label{ampNLO} \\
& +\left( {\frac{g_{A}^{2}}{2f^{2}}}\right) \left( -1+{\frac{m_{\pi }^{2}}{%
4p^{2}}}\ln \left( 1+{\frac{4p^{2}}{m_{\pi }^{2}}}\right) \right)  \notag \\
& +{\frac{g_{A}^{2}}{f^{2}}}\left( {\frac{m_{\pi }M\mathcal{A}_{-1}}{4\pi }}%
\right) \left\{ -{\frac{(\mu +ip)}{m_{\pi }}}+{\frac{m_{\pi }}{2p}}\left[
\tan ^{-1}\left( {\frac{2p}{m_{\pi }}}\right) +{\frac{i}{2}}\ln \left( 1+{%
\frac{4p^{2}}{m_{\pi }^{2}}}\right) \right] \right\}  \notag \\
& +{\frac{g_{A}^{2}}{2f^{2}}}\left( {\frac{m_{\pi }M\mathcal{A}_{-1}}{4\pi }}%
\right) ^{2}\left\{ 1-\left( {\frac{\mu +ip}{m_{\pi }}}\right) ^{2}+i\tan
^{-1}\left( {\frac{2p}{m_{\pi }}}\right) -{\frac{1}{2}}\ln \left( {\frac{%
m_{\pi }^{2}+4p^{2}}{\mu ^{2}}}\right) \right\} .  \notag
\end{align}%
where $g_{A}$ is the pion nucleon coupling constant, $f$ is the pion decay
constant, and we have imposed isospin symmetry by setting $m_{u}=m_{d}$ and
neglecting the electromagnetic interaction. $C_{0,0}$ is a NLO operator with
the same structure as $C_{0}$. $D_{2}$ is a non-derivative four-nucleon
coupling with one insertion of $m_{q}$ (or $m_{\pi }^{2}$), and $C_{2}$ is a
two-derivative four-nucleon operator that is independent of $m_{q}$. $\mu $
is the renormalization scale, and we have used dimensional regularization
and the power-divergence subtraction procedure (PDS)~\cite{Kaplan:1998we} to
renormalize the theory. These amplitudes are manifestly renormalization
scale independent order-by-order in the EFT expansion.

Expanding the right hand side of Eq.(\ref{eq1}) in powers of $p$, we have
the matching for the spin singlet and triplet scattering lengths

\begin{eqnarray}
\frac{1}{a^{({}^{1}S_{0})}} &=&\gamma ^{({}^{1}S_{0})}-\frac{M}{4\pi }%
(\gamma ^{({}^{1}S_{0})}-\mu )^{2}\ \left( D_{2}^{(^{1}S_{0})}m_{\pi
}^{2}+C_{0,0}^{(^{1}S_{0})}\right)  \notag \\
&&+\frac{g_{A}^{2}M}{8\pi f^{2}}\left[ m_{\pi }^{2}\,\log {\left( {\frac{\mu 
}{m_{\pi }}}\right) }+(\gamma ^{({}^{1}S_{0})}-m_{\pi })^{2}-(\gamma
^{({}^{1}S_{0})}-\mu )^{2}\right]  \notag \\
\frac{1}{a^{({}^{3}S_{1})}} &=&\gamma ^{({}^{3}S_{1})}-\frac{M}{4\pi }%
(\gamma ^{({}^{3}S_{1})}-\mu )^{2}\ \left[ D_{2}^{({}^{3}S_{1})}\left(
m_{\pi }^{2}-\epsilon \lambda ^{2}\right) +C_{0,0}^{({}^{3}S_{1})}\right] 
\notag \\
&&+\frac{g_{A}^{2}M}{8\pi f^{2}}\left[ m_{\pi }^{2}\,\log {\left( {\frac{\mu 
}{m_{\pi }}}\right) }+(\gamma ^{({}^{3}S_{1})}-m_{\pi })^{2}-(\gamma
^{({}^{3}S_{1})}-\mu )^{2}\right]  \notag \\
&&-\epsilon \frac{g_{A}^{2}M}{8\pi f^{2}}\left[ \lambda ^{2}\,\log {\left( {%
\frac{\mu }{\lambda }}\right) }+(\gamma ^{({}^{3}S_{1})}-\lambda
)^{2}-(\gamma ^{({}^{3}S_{1})}-\mu )^{2}\right] ,  \label{a}
\end{eqnarray}%
where $\gamma ^{(i)}=\mu +4\pi /MC_{0}^{(i)}$ is the LO inverse scattering
length. We perform the expansion around the physical pion mass $m_{\pi
}^{phys}\simeq 138$ MeV, so $\gamma ^{(i)}$ takes the physical value $%
1/a_{phys}^{(i)}$. To fix $D_{2}^{(i)}$ and $C_{0,0}^{(i)}$, we just need $%
1/a^{(i)}$ computed at another $m_{\pi }$ other than $m_{\pi }^{phys}$.

The matching for effective ranges gives 
\begin{eqnarray}
r_{0}^{({}^{1}S_{0})} &=&\frac{MC_{2}^{({}^{1}S_{0})}(\mu -\gamma
^{({}^{1}S_{0})})^{2}}{2\pi }+\frac{g_{A}^{2}M}{12\pi f^{2}}\left[ 6\left( 
\frac{\gamma ^{({}^{1}S_{0})}}{m_{\pi }}\right) ^{2}-8\frac{\gamma
^{({}^{1}S_{0})}}{m_{\pi }}+3\right] \   \notag \\
r_{0}^{({}^{3}S_{1})} &=&\frac{MC_{2}^{({}{}^{3}S_{1})}(\mu -\gamma
^{({}{}^{3}S_{1})})^{2}}{2\pi }  \notag \\
&&+\frac{g_{A}^{2}M}{12\pi f^{2}}\left[ 6\left( \frac{\gamma
^{({}{}^{3}S_{1})}}{m_{\pi }}\right) ^{2}-8\frac{\gamma ^{({}{}^{3}S_{1})}}{%
m_{\pi }}+3-\epsilon \left[ 6\left( \frac{\gamma ^{({}{}^{3}S_{1})}}{\lambda 
}\right) ^{2}-8\frac{\gamma ^{({}{}^{3}S_{1})}}{\lambda }+3\right] \right] .
\label{r0}
\end{eqnarray}%
Unlike the scattering lengths, no lattice data is needed to study the quark
mass dependence of the effective ranges since $C_{2}^{(i)}$ can be fixed by $%
r_{0}^{({}i)}$ at $m_{\pi }^{phys}$.

Note that the $\epsilon $ terms in $1/a^{({}^{3}S_{1})}$ and $%
r_{0}^{({}^{3}S_{1})}$ are $m_{\pi }$ independent, so they can be absorbed
into counterterms $C_{0,0}^{({}^{3}S_{1})}$ and $C_{2}^{({}{}^{3}S_{1})}$.
Therefore, the KSW and BKV schemes give the same $m_{\pi }$ dependence to
ERP's at NLO.

In summary, Eqs.(\ref{a},\ref{r0}) can be parametrized as%
\begin{eqnarray}
\frac{1}{a^{(i)}} &=&\overline{\gamma }^{({}i)}-d_{2}^{(i)}m_{\pi }^{2}+%
\frac{g_{A}^{2}M}{8\pi f^{2}}\left[ m_{\pi }^{2}\,\log {\left( {\frac{\mu }{%
m_{\pi }}}\right) }+(\gamma ^{({}i)}-m_{\pi })^{2}\right]  \notag \\
r_{0}^{({}i)} &=&c_{2}^{({}i)}+\frac{g_{A}^{2}M}{12\pi f^{2}}\left[ 6\left( 
\frac{\gamma ^{({}i)}}{m_{\pi }}\right) ^{2}-8\frac{\gamma ^{({}i)}}{m_{\pi }%
}\right] .
\end{eqnarray}%
The physical $a^{(i)}$ and $r_{0}^{({}i)}$ ($a_{phys}^{({}^{3}S_{1})}=5.423%
\pm 0.005$ fm, $r_{0,phys}^{({}^{3}S_{1})}=1.764\pm 0.002$ fm, $%
a_{phys}^{({}^{1}S_{0})}=-23.714\pm 0.003$ fm, $%
r_{0,phys}^{({}^{1}S_{0})}=2.73\pm 0.03$ fm) fix $c_{2}^{({}i)}$ and a
combination of $\overline{\gamma }^{({}i)}$ and $d_{2}^{(i)}$. We only need
a LQCD calculation of $a^{(i)}$ at different $m_{\pi }$ to get the leading $%
m_{\pi }$ dependence for $a^{(i)}$ and $r_{0}^{({}i)}$.


\begin{figure}[tbp]
\begin{center}
\includegraphics[height=5cm]{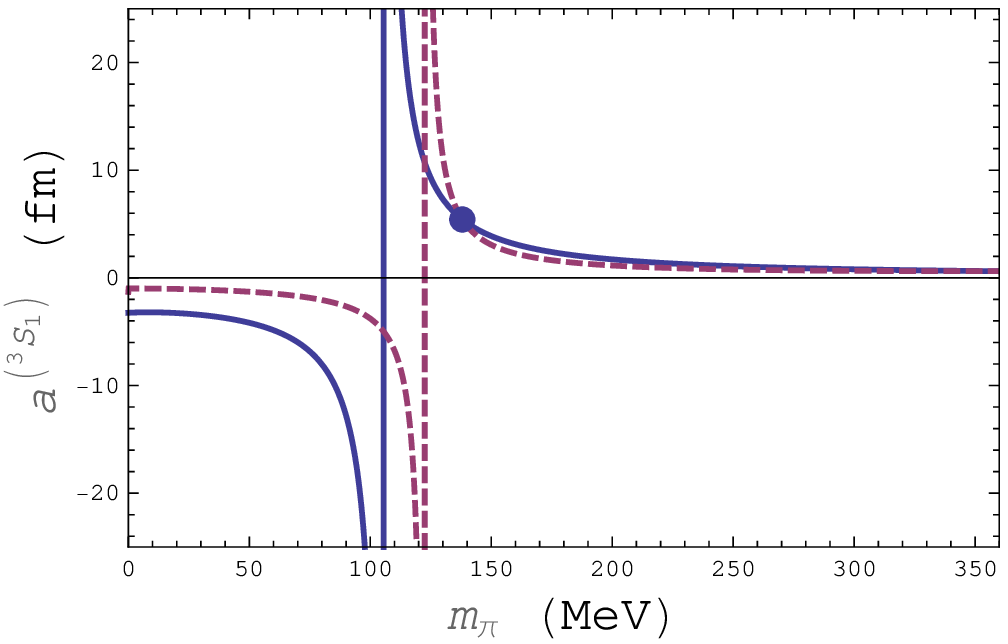} \includegraphics[height=5cm]{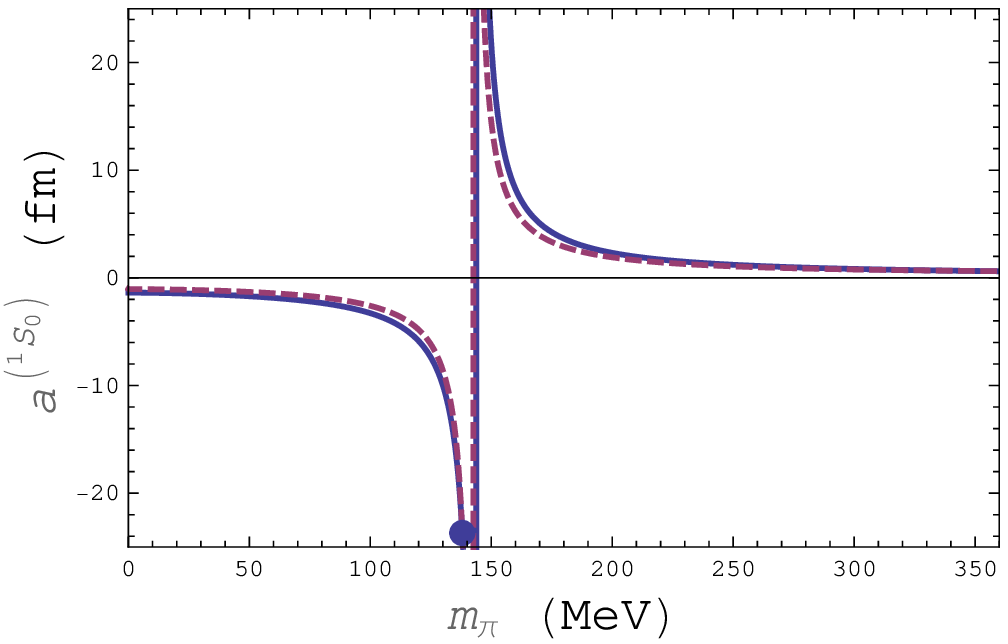}
\end{center}
\caption{Scattering lengths of the ${}^{3}S_{1}$ and ${}^{1}S_{0}$ states
vs. $m_{\protect\pi }$ using the NLO BKV result of Eq.(\protect\ref{a}), the
physical scattering length, and the scattering length computed at $m_{%
\protect\pi }=353$ MeV with lattice QCD. The dashed(solid) lines are
with(without) the higher order $m_{\protect\pi }$ dependence in $M$, $f$ and 
$g_{A}$ included. The dot is the physical point.}
\end{figure}


\begin{figure}[tbp]
\begin{center}
\includegraphics[height=5cm]{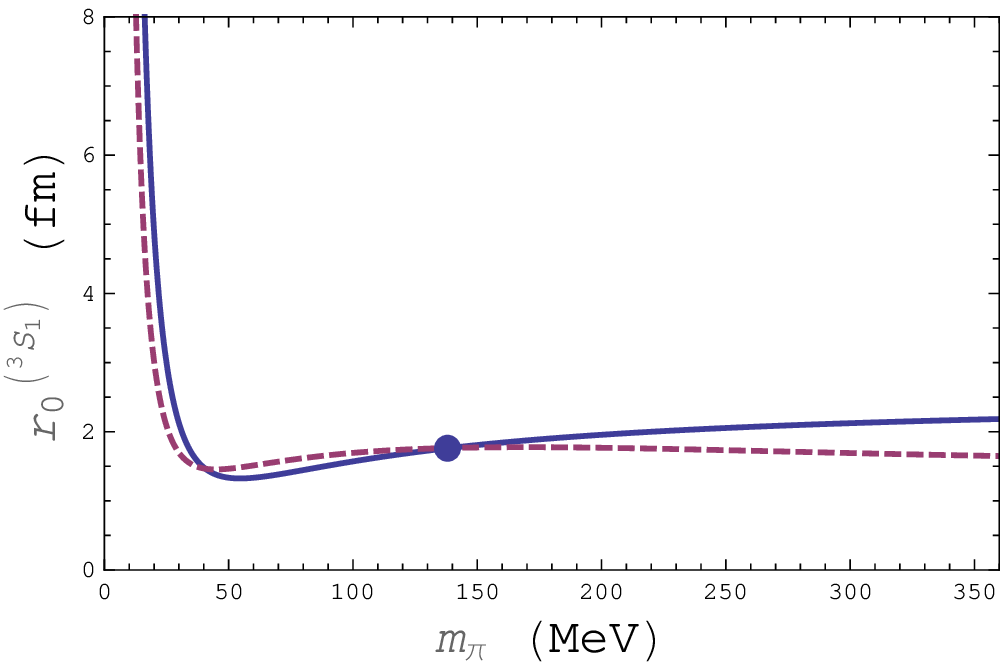} \includegraphics[height=5cm]{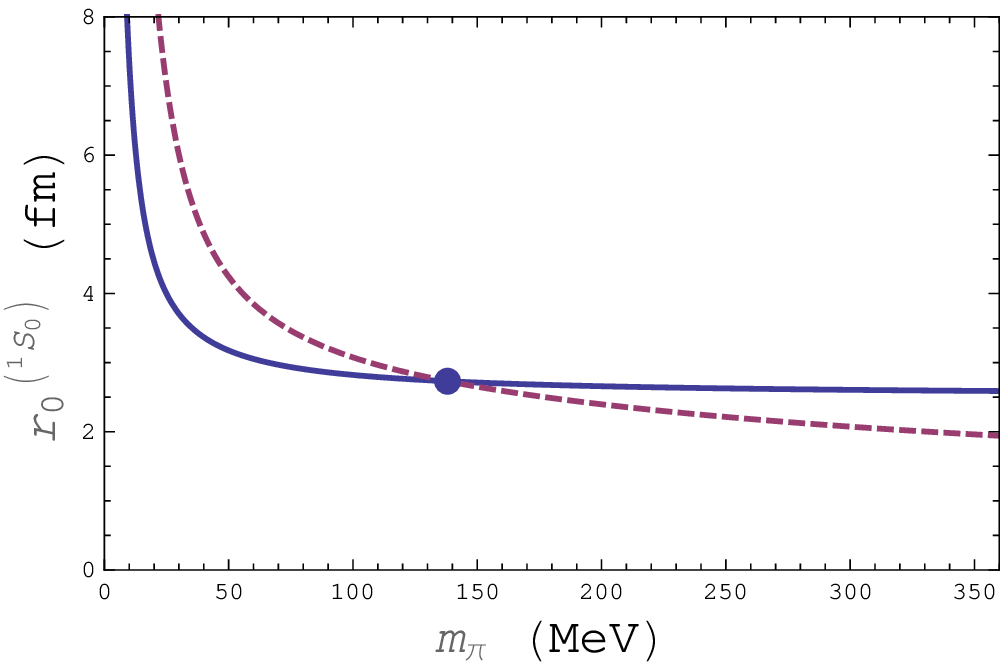}
\end{center}
\caption{Effective ranges of the ${}^{3}S_{1}$ and ${}^{1}S_{0}$ states vs. $%
m_{\protect\pi }$ using the NLO BKV result of Eq.(\protect\ref{r0}). The
notations are the same as in Fig. 1.}
\end{figure}

\begin{figure}[tbp]
\begin{center}
\includegraphics[height=5cm]{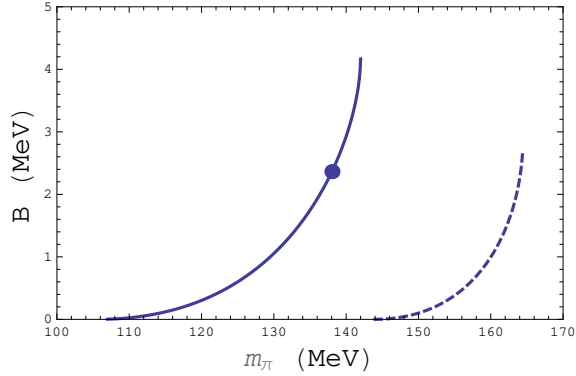}
\end{center}
\caption{Bound state (whenever exists) binding energy vs. $m_{\protect\pi }$
for $^{3}S_{1}$(solid line) and ${}^{1}S_{0}$ (dashed line). The notations
are the same as in Fig. 1.}
\end{figure}

Currently, the smallest $m_{\pi }$ that $a^{(i)}$ is computed on the lattice
is $353.7\pm 2.1$ MeV \cite{Beane:2006mx}. The calculation yields $%
a^{({}^{3}S_{1})}=0.63\pm 0.74$ fm, $a^{({}^{1}S_{0})}=0.63\pm 0.50$ fm. The
central values yield the solid curves in Fig. 1 and 2. We can study the size
of higher order corrections by including the $m_{\pi }$ dependence of $M$, $%
f $ and $g_{A}$ (these are next-to-next-to-leading-order corrections) which
is extracted from lattice data \cite{Edwards:2005ym,Beane:2005rj} to Eqs.(%
\ref{a},\ref{r0}). This yields the dashed curves in Fig. 1 and 2. The $%
a^{({}^{3}S_{1})}\rightarrow \infty $ position can shift by $\sim 20\%$ in $%
m_{\pi }$\ due to higher order corrections, while the corrections to $%
a^{({}^{1}S_{0})}$ is much smaller. When $m_{\pi }\gtrsim 100$ MeV, $%
r_{0}^{({}i)}\simeq 2$ fm and is insensitive to $m_{\pi }$.

The analytic structure of the scattering amplitude, Eq.(\ref{A}), is that
there are two cuts from $p=im_{\pi }/2$ to $i\infty $ and from $p=-im_{\pi
}/2$ to $-i\infty $. There is a ${}^{3}S_{1}$ bound state for $m_{\pi }=106$
to $142$ MeV (with $\lambda =750$ MeV, but the range remains the same for $%
\lambda =500$ to $1000$ MeV) and a ${}^{1}S_{0}$ bound state for $m_{\pi
}=144$ to $165$ MeV. The corresponding binding energies are shown in Fig. 3.
This result can be understood by examining the scattering amplitude in the
effective range expansion. By keeping only the scattering length and
effective range in Eq.(\ref{delta}), the amplitude of Eq.(\ref{A}) has two
poles 
\begin{equation}
p=\frac{i}{r_{0}}\left( 1\pm \sqrt{1-\frac{2r_{0}}{a}}\right) .  \label{p}
\end{equation}%
If $a>0$, the solution with smaller $\left\vert p\right\vert $ is $p=\frac{i%
}{r_{0}}\left( 1-\sqrt{1-\frac{2r_{0}}{a}}\right) $. The bound state exists
when $0<\frac{2r_{0}}{a}<1$. On the other hand, if $a<0$, the solution with
smaller $\left\vert p\right\vert $ is $p=\frac{i}{r_{0}}\left( 1-\sqrt{1-%
\frac{2r_{0}}{a}}\right) $. Since $\frac{2r_{0}}{a}<0$, the pole does not
correspond to a bound state. The other pole $p=\frac{i}{r_{0}}\left( 1+\sqrt{%
1-\frac{2r_{0}}{a}}\right) $ (for both positive and negative $a$) is of the
order of the ultraviolet cut-off scale $1/r_{0}$ which is usually hidden in
the cut starting at $p=im_{\pi }/2$. Thus, the bound state range is $%
0<2r_{0}<a$, which is close to the ranges seen in Fig. 3. Furthermore, the
maximum binding momentum is $i/r_{0}$, or the maximum binding energy is $%
1/(Mr_{0}^{2})$ $\sim 6$ MeV for $r_{0}\sim 2.5$ fm.

With the current lattice input, this theory does not have a two nucleon
bound state in the chiral ($m_{q}\rightarrow 0$, or equivalently $m_{\pi
}\rightarrow 0$) limit. However, different power countings could lead to
different conclusions \cite{Beane:2006mx}. This might indicate that $m_{\pi
}=354$ MeV is not within the common \textquotedblleft chiral
regime\textquotedblright , i.e. within the radius of convergence of the $%
m_{\pi }$ expansion, for these theories. It is important to perform higher
order EFT calculations to decide the size of the chiral regime and to answer
how small $m_{\pi }$ should be for future LQCD calculations to draw a firm
conclusion about the deuteron binding energy in the chiral limit.

\subsection{A support for multiverse?}

It is curious that $m_{\pi }^{phys}$ is so close to the upper bound of $%
m_{\pi }$ where the deuteron is bounded---if $m_{\pi }^{phys}$ were $5\%$
bigger, then there would not have been deuteron at all. This makes it much
harder for primordial nuclear synthesis to form light nuclei through the
usual pathways and might eventually make life impossible. This interesting
fine tuning implies that our universe sits near the edge of the parameter
space where life could exist. In Ref. \cite{Bousso:2009ks}, it is argued
that this is not a fine tuning but a natural case if multiverse exists: In a
multiverse, which is an ensemble of many universes including ours, the
majority of the universes do not allow life to exist since it requires lots
of conditions to be satisfied. Thus, the peak of the $m_{q}$\ distribution
in this multiverse will be more likely to sit outside the parameter space
where life is possible. In that case, the tail of the distribution goes
across this parameter space and then one finds that most of the universes
that permits life is near the edge of the parameter space. Thus, if the
multiverse exists, without fine tuning, our universe should live near the
edge of the parameter space where life is possible (called the catastrophic
boundary in \cite{Bousso:2009ks}). It is interesting to note that our case
of the deuteron bound state is consistent with this pattern, similar to the
example of the cosmological constant whose value is close to the allowed
range obtained by Weinberg through the anthropic principle \cite%
{Weinberg:1987dv,Martel:1997vi} and several examples worked out in \cite%
{Bousso:2009ks}. Although we do not consider this as a sharp test of the
multiverse conjecture, because the conjecture cannot be falsified even if
the physical $m_{q}$ is far away from the edge of the allowed parameter
space, it is still interesting to see whether there are cases being
consistent with this conjecture.

\section{Matching between the Theory with and without Pions}

We are interested in using the theory without pions to describe low energy
processes (where $p<m_{\pi }$ so pions can be integrated out) at
non-physical $m_{\pi }$. The matching between the pionful and pionless EFT's
at those $m_{\pi }$ gives the $m_{\pi }$ dependence of the couplings in the
pionless theory. Those couplings are the ERP's of NN scattering mentioned
above and current operators when coupled to external currents.

We can classify the non-derivative single-nucleon (one-body) current
operators by how they transform in the spin-isospin space: the scalar-scalar
operator ($N^{\dagger }N$), scalar-vector operator ($N^{\dagger }\tau _{i}N$%
), vector-scalar operator ($N^{\dagger }\sigma _{i}N$) and vector-vector
operator ($N^{\dagger }\sigma _{i}\tau _{j}N$), where $\sigma _{i}$($\tau
_{i}$) acts on the spin(isospin) space and the spacial indexes $i,j=1,2,3$.
The non-derivative scalar-scalar and scalar-vector operators originate from
matrix elements of the quark level operators $\overline{q}\gamma _{0}q$ and $%
\overline{q}\gamma _{0}\tau _{i}q$. They do not have two-body currents due
to vector current conservation. For vector-scalar currents, they could
originate from matrix elements of the isoscalar quark axial operator $%
\overline{q}\gamma _{i}\gamma _{5}q$ or the magnetic part of the vector
current $\overline{q}\gamma _{i}q$, so the corresponding two body-currents
exist. For vector-vector currents, the corresponding quark level operator is 
$\overline{q}\gamma _{i}\gamma _{5}\tau _{j}q$ and the two body-currents
(called Gamow-Teller operators) also exist.

From matching the isoscalar magnetic current between the theory with and
without pions, we conclude that the vector-scalar two-body currents do not
depend on pion mass at the leading order \cite{Kaplan:1998sz}. The matching
of the two-body Gamow-Teller operator \cite{Butler:1999sv,Butler:2000zp}
yields 
\begin{eqnarray}
L_{GT} &=&l_{GT}-\frac{\kappa _{1}g_{A}^{2}m_{\pi }^{2}}{2\gamma ^{^{2}}f^{2}%
}\,\log {\left( {\frac{m_{\pi }}{m_{\pi }+2\gamma }}\right) }  \notag \\
&&{-}\frac{\kappa _{1}g_{A}^{2}}{6a\gamma f^{2}m_{\pi }^{2}\left( m_{\pi
}+2\gamma \right) }\left[ 6a^{({}^{1}S_{0})}m_{\pi }^{4}+m_{\pi }^{2}\left(
9a^{({}^{1}S_{0})}m_{\pi }-4\right) \gamma -2m_{\pi }\gamma ^{2}\left(
a^{({}^{1}S_{0})}m_{\pi }-5\right) \right.   \notag \\
&&\left. -2\gamma ^{3}\left( 5a^{({}^{1}S_{0})}m_{\pi }+6\right)
+12a^{({}^{1}S_{0})}\gamma ^{4}\right] ,  \label{GT}
\end{eqnarray}%
where $l_{GT}$ is $m_{\pi }$ independent, $\gamma =\left( 1-\sqrt{%
1-2r_{0}^{(^{3}S_{1})}/a^{(^{3}S_{1})}}\right) /r_{0}^{(^{3}S_{1})}$ is the
deuteron binding momentum and $\kappa _{1}$ is the single nucleon coupling
(for the isovector magnetic current, $\kappa _{1}$ is the isovector nucleon
magnetic moment; for weak coupling, $\kappa _{1}$ is proportional to $g_{A}$%
). There is no unknown parameter in the $m_{\pi }$ dependent term.

\section{The Quark Mass Dependence of More Two Nucleon Observables}

In this section we apply the $m_{\pi }$ dependent couplings in the pionless
EFT, which has been worked out in the previous sections, to compute several
physical observables involving deuterons.

\begin{figure}[tbp]
\begin{center}
\includegraphics[height=5cm]{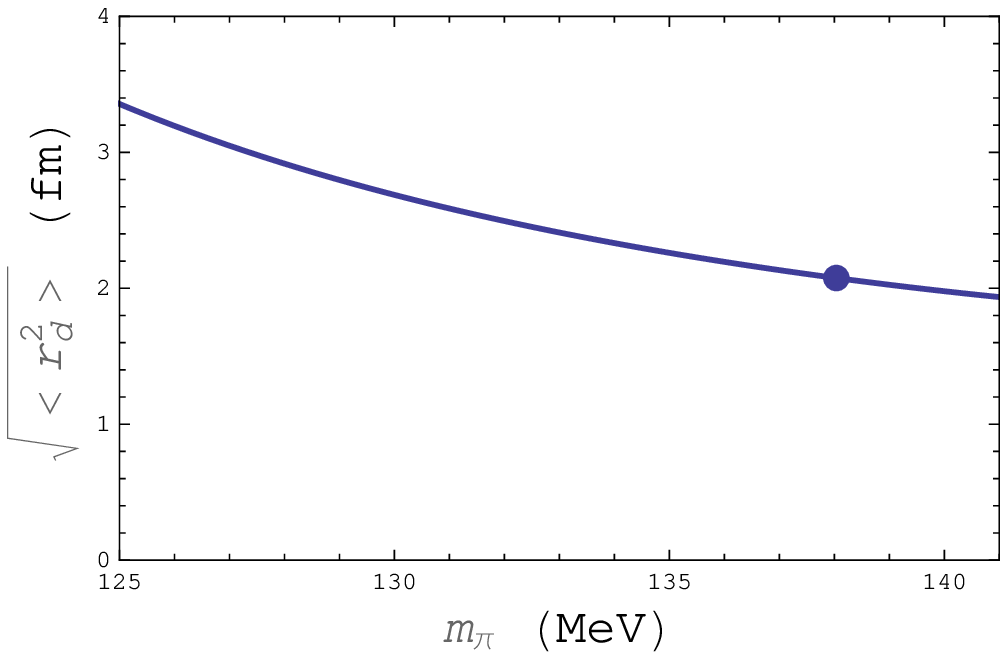} %
\includegraphics[height=5cm]{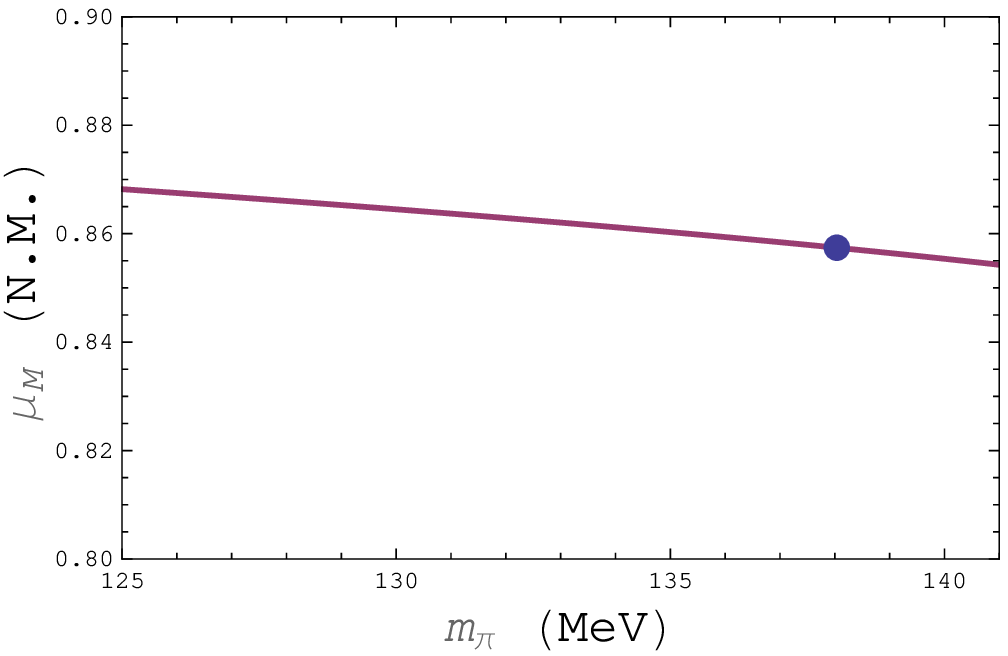} \includegraphics[height=5cm]{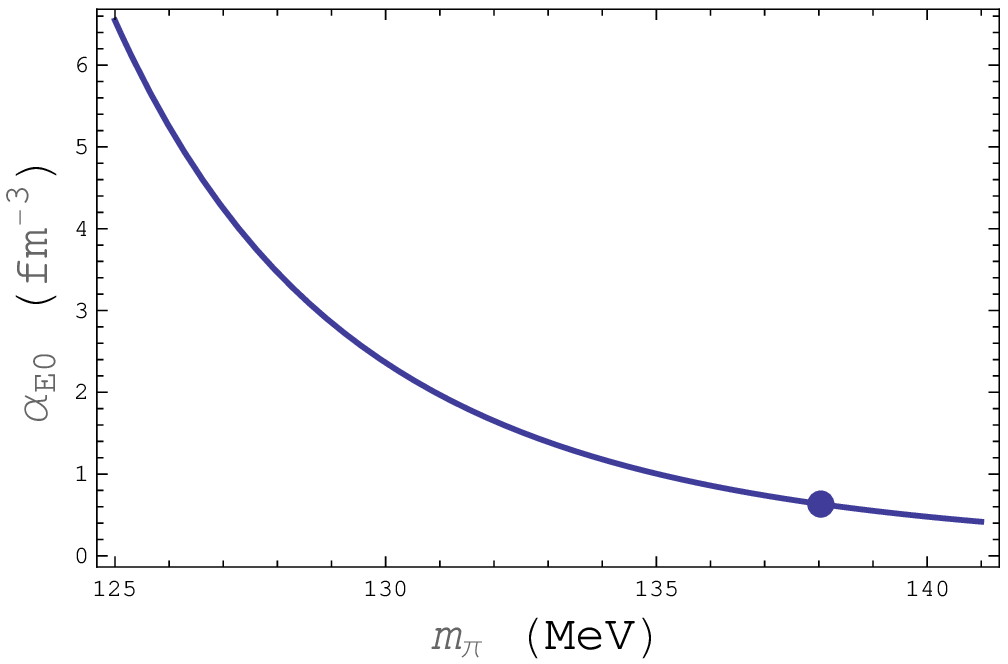}
\end{center}
\caption{Deuteron charge radius ($\protect\sqrt{\left\langle
r_{d}^{2}\right\rangle }$), magnetic moment ($\protect\mu _{M}$), and
electric polarizability ($\protect\alpha _{E0}$) vs. $m_{\protect\pi }$. The
dots are the physical points.}
\end{figure}

\subsection{Deuteron properties}

The deuteron charge radius has the expression \cite{Chen:1999tn} 
\begin{equation}
\left\langle r_{d}^{2}\right\rangle =\left\langle r_{N,0}^{2}\right\rangle +%
\frac{1}{8\gamma ^{2}\left( 1-\gamma \rho _{d}\right) }
\end{equation}%
where the isoscalar charge radius of the nucleon $\sqrt{\left\langle
r_{N,0}^{2}\right\rangle }=0.79\pm 0.01$ fm and $\rho
_{d}=r_{0}^{({}^{3}S_{1})}$. As expected, the deuteron charge radius$\sqrt{%
\left\langle r_{d}^{2}\right\rangle }$ is set by the inverse binding
momentum $1/\gamma $ when the nucleon charge radius is negligible. The $%
m_{\pi }$ dependence for $m_{\pi }=125-141$ MeV (where deuteron is bounded)
is shown in Fig. 4.


The deuteron magnetic moment is \cite{Chen:1999tn} 
\begin{equation}
\mu _{M}=\frac{e}{2M}\left( 2\kappa _{0}+\gamma L_{V-S}\right) ,
\end{equation}%
where $\kappa _{0}=0.44$ is the nucleon isoscalar magnetic moment in units
of nuclear magneton (N.M.) and the vector-scalar two-body current $L_{V-S}$
is $m_{\pi }$ independent. Neither the one-nucleon nor the two-nucleon
contribution is sensitive to $m_{\pi }$. The sum is also shown in Fig. 4.

The deuteron polarizability is computed as \cite{Chen:1999tn} 
\begin{equation}
\alpha _{E,0}=\frac{\alpha M}{32\gamma ^{4}\left( 1-\gamma \rho _{d}\right) }%
,
\end{equation}%
$\alpha =1/137$ is the fine structure constant. It has a strong $m_{\pi }$
dependence\ as is shown in Fig. 4.

\subsection{Reaction: $np\leftrightarrow d\protect\gamma $}

The process $np\leftrightarrow d\gamma $ is relevant for BBN. Its cross
section is proportional to the wave function overlap between the initial and
final states. Since the deuteron size $1/\gamma $ is very sensitive to $%
m_{\pi }$ near $m_{\pi }^{phys}$, the cross section also changes
dramatically in this region.


\begin{figure}[tbp]
\begin{center}
\includegraphics[height=7cm]{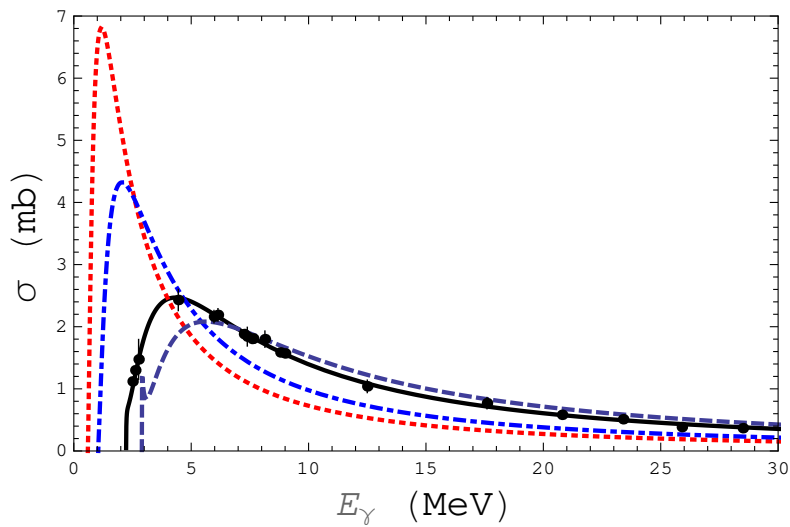}
\end{center}
\caption{The cross section for $\protect\gamma d\rightarrow np$ as a
function of the incident photon energy in MeV. The solid curve is for the
physical $m_{\protect\pi }$, while the dotted, dotdashed, and dashed curves
are for $m_{\protect\pi }=$ 125, 130, and 141 MeV, respectively.}
\end{figure}

The total cross section for $np\rightarrow d\gamma $ is \cite%
{Chen:1999bg,Rupak:1999rk} 
\begin{equation*}
\sigma \left( np\rightarrow d\gamma \right) ={\frac{4\pi \alpha \left(
\gamma ^{2}+p^{2}\right) ^{3}}{\gamma ^{3}M^{4}p}}\left[ \ |\tilde{X}%
_{M1}|^{2}\ +\ |\tilde{X}_{E1}|^{2}\ \right] \ \ \ ,
\end{equation*}%
where $p$ is the magnitude of the momentum of each nucleon in the
center-of-mass frame. The electric dipole ($E1$) transition yields 
\begin{equation}
|\tilde{X}_{E1}|^{2}={\frac{p^{2}M^{2}\gamma ^{4}}{\left( \gamma
^{2}+p^{2}\right) ^{4}}}\left[ 1+\gamma \rho _{d}+(\gamma \rho
_{d})^{2}+\cdots \right] \ \ \ .  \label{eq:Eone}
\end{equation}%
The Magnetic dipole ($M1$) transition yields 
\begin{equation*}
|\tilde{X}_{M1}|^{2}={\frac{\kappa _{1}^{2}\gamma ^{4}\left( {\frac{1}{%
a^{({}^{1}S_{0})}}}-\gamma \right) ^{2}}{\left( {\frac{1}{%
a^{({}^{1}S_{0})^{2}}}}+p^{2}\right) \left( \gamma ^{2}+p^{2}\right) ^{2}}}%
\left[ 1+\gamma \rho _{d}-r_{0}^{({}^{1}S_{0})}{\frac{\left( {\frac{\gamma }{%
a^{({}^{1}S_{0})}}}+p^{2}\right) p^{2}}{\left( {\frac{1}{%
a^{({}^{1}S_{0})^{2}}}}+p^{2}\right) \left( {\frac{1}{a^{({}^{1}S_{0})}}}%
-\gamma \right) }}-{\frac{L_{GT}}{\kappa _{1}}}{\frac{M}{2\pi }}{\frac{%
\gamma ^{2}+p^{2}}{{\frac{1}{a^{({}^{1}S_{0})}}}-\gamma }}\right] \ \ \ .
\end{equation*}%
The Gamow-Teller two-body current: $L_{GT}=-4.513~\mathrm{fm}^{2}$ at $%
m_{\pi }=m_{\pi }^{phys}$, is fitted from the measured cross section $\sigma
^{\mathrm{expt}}=334.2\pm 0.5~\mathrm{mb}$ \cite{Chen:1999bg} using incident
neutrons of speed $|v|=2200~\mathrm{m/s}$. The $m_{\pi }$ dependence of $%
L_{GT}$ is shown in Eq.(\ref{delta}). The isovector nucleon magnetic moment $%
\kappa _{1}=2.35-g_{A}^{2}M(m_{\pi }-m_{\pi }^{phys})/\left( 2\pi
f^{2}\right) $, where we have applied the $m_{\pi }$ dependence calculated
from\ ChPT \cite{Bernard:1995dp}. The cross section of the reverse process
with deuteron being at rest is%
\begin{equation}
\sigma \left( \gamma d\rightarrow np\right) =\frac{2M\left( E_{\gamma
}-B\right) }{3E_{\gamma }^{2}}\sigma \left( np\rightarrow d\gamma \right) ,
\end{equation}%
where $E_{\gamma }$ is the incident photon energy. This deuteron
photo-disintegration cross section for $m_{\pi }=125-141$ MeV is shown in
Fig. 5.

\section{Conclusion}

We have studied the implications of lattice QCD determinations of the S-wave
nucleon-nucleon scattering lengths at unphysical light quark masses. It is
found that with the help of nuclear effective field theory, not only the
quark mass dependence of the effective range parameters, but also the
leading quark mass dependence of all the low energy (with $p\ll m_{\pi }$)
deuteron matrix elements can be obtained. The quark mass dependence of
deuteron charge radius, magnetic moment, polarizability and the deuteron
photodisintegration cross section are shown based on the NPLQCD lattice
calculation of the scattering lengths at 354 MeV pion mass and the NEFT
power counting scheme of Beane, Kaplan and Vuorinen. Further improvement can
be obtained by performing the lattice calculation at smaller quark masses.
But at the same time, it is important to perform higher order EFT
calculations to decide the radius of convergence in $m_{\pi }$ in order to
answer how small $m_{\pi }$ should be for future LQCD calculations to
provide reliable $m_{\pi }$ dependence for two nucleon observables all the
way to the chiral limit.

Our result can be used to constrain the time variation of isoscalar
combination of $u$ and $d$ quark mass $m_{q}$, to help the anthropic
principle study to find the $m_{q}$ range which allows the existence of
life, and to provide a weak test of the multiverse conjecture. \ 

\bigskip

We thank Martin Savage for providing us the formulae describing the LQCD $%
m_{\pi }$ dependence of $M$, $f$ and $g_{A}$. This work is supported by the
NSC and NCTS of ROC.

\end{document}